# Nuclear quantum effects in molecular dynamics simulations

**H Dammak**[1], **M Hayoun**[2], **F Brieuc**[1], **G Geneste**[3]

[1] Laboratoire Structures Propriétés et Modélisation des Solides, CentraleSupélec, CNRS, Université Paris-Saclay, 91190 Gif-sur-Yvette, France
[2] Laboratoire des Solides Irradiés, Ecole Polytechnique, CNRS, CEA, Université Paris-Saclay, 91128 Palaiseau, France
[3] CEA, DAM, DIF, F-91297 Arpajon, France

hichem.dammak@centralesupelec.fr

**Abstract**. To take into account nuclear quantum effects on the dynamics of atoms, the path integral molecular dynamics (PIMD) method used since 1980s is based on the formalism developed by R. P. Feynman. However, the huge computation time required for the PIMD reduces its range of applicability. Another drawback is the requirement of additional techniques to access time correlation functions (ring polymer MD or centroid MD). We developed an alternative technique based on a quantum thermal bath (QTB) which reduces the computation time by a factor of ~20. The QTB approach consists in a classical Langevin dynamics in which the white noise random force is replaced by a Gaussian random force having the power spectral density given by the quantum fluctuation-dissipation theorem. The method has yielded satisfactory results for weakly anharmonic systems: the quantum harmonic oscillator, the heat capacity of a MgO crystal, and isotope effects in $^7$LiH and $^7$LiD. Unfortunately, the QTB is subject to the problem of zero-point energy leakage (ZPEL) in highly anharmonic systems, which is inherent in the use of classical mechanics. Indeed, a part of the energy of the high-frequency modes is transferred to the low-frequency modes leading to a wrong energy distribution. We have shown that in order to reduce or even eliminate ZPEL, it is sufficient to increase the value of the frictional coefficient. Another way to solve the ZPEL problem is to combine the QTB and PIMD techniques. It requires the modification of the power spectral density of the random force within the QTB. This combination can also be seen as a way to speed up the PIMD.

## 1. Introduction

Molecular dynamics (MD) simulation is frequently used to investigate and predict the properties of condensed matter in the classical limit. For a crystal, these calculations are valid for temperatures higher than the Debye temperature [1] (940 K for MgO). As shown in figure 1, the experimental heat capacity decreases when the temperature decreases and vanishes at $T = 0$ K, while MD simulation leads to a nearly constant value, corresponding to $3k_B$ per atom. The arrow in figure 1 points out this disagreement. Standard MD ignores the nuclear quantum effects (NQE), which are at the origin of the shape of the experimental heat capacity curve. This behavior is a direct consequence of the quantization of the energy of the vibration modes. Another example is the evolution as a function of temperature of the lattice parameter in lithium hydride (LiH) given in figure 2. At low temperature, the crystal rapidly reaches its ground state upon cooling, generating a freezing (saturation) of the lattice parameter whereas the standard MD predicts a linear behavior. The arrow in figure 2 emphasizes this discrepancy due to the quantization of the energy of the vibration modes and due to the total energy of the ground state, called zero-point energy (ZPE), which is larger than the potential energy minimum. In addition, the lattice parameter of lithium deuteride (LiD) behaves similarly but is shifted (see double arrow) with respect to the LiH one. This isotope effect, which is a particular case of NQE, is due to the change of the vibrational frequencies, in turn due to the difference in the atomic masses of H and D. In fact, the average energy of the quantum oscillator, at a given temperature, is frequency dependent. In the harmonic approximation, the average energy is given by the well-known expression:

$$\theta(\omega, T) = \hbar\omega \left[\frac{1}{2} + \frac{1}{\exp(\beta\hbar\omega) - 1}\right] \qquad (1)$$

where $\omega$ is the angular frequency of the oscillator, $\hbar$ is the reduced Planck constant and $\beta$ the statistical temperature ($1/k_B T$). Figure 3 shows the evolution of this energy as a function of temperature compared to the classical limit ($k_B T$). One can note the freezing of the energy at low temperature ($T < \hbar\omega/k_B$) corresponding to the ZPE, $\hbar\omega/2$. In contrast, in the classical description, the average energy does not depend on the frequency of the oscillator.

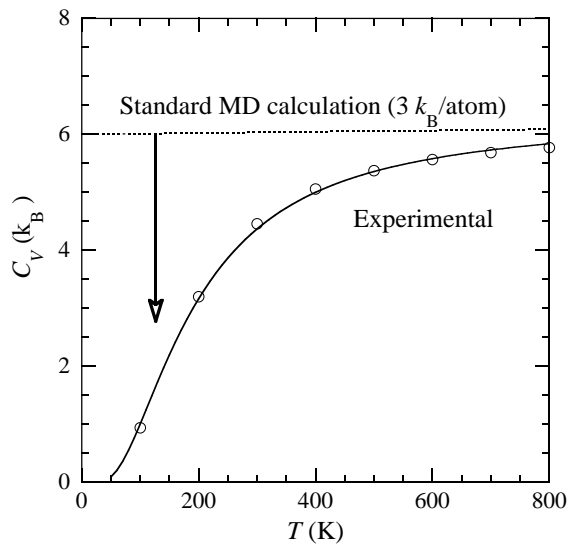
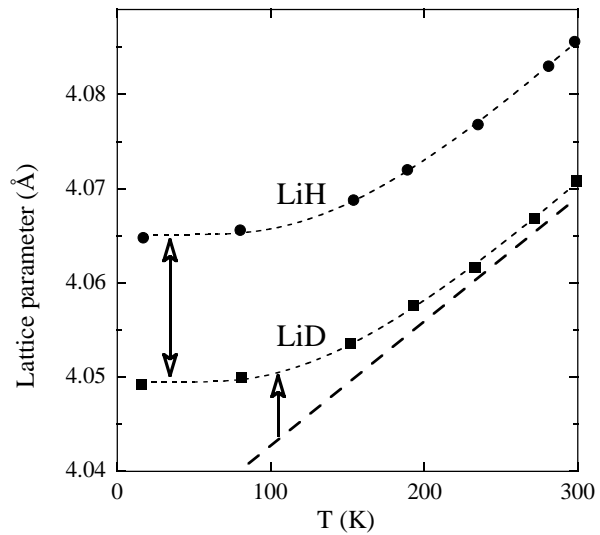

**Figure 1.** Heat capacity per molecule of the MgO crystal as a function of the temperature, $T$ [2]. The full line is the Debye model [1].

**Figure 2.** Experimental lattice parameters of LiH and LiD crystals as a function of the temperature, $T$ [3]. The straight dashed line is the standard MD result.

These examples show that the quantum nature of nuclei can play a major role at low temperatures and/or in systems that contain light atoms. In this case, NQE cannot be neglected and must be taken into account especially in MD simulation, which is the focus of this paper. The historical method including NQE is based on Feynman's path integral [4] (PIMD or path integral Monte Carlo). It can provide exact results − under the assumption of distinguishable nuclei − but at the price of a high computational cost. An (approximate) alternative method includes the quantum fluctuations through a random force, whose power spectral density is related to $\theta(\omega, T)$ by the quantum mechanical fluctuation-dissipation theorem [5]. This is the quantum thermal bath (QTB) MD introduced in 2009 by Dammak *et al.* [6]. The two methods can be combined, as QTB-PIMD [7], to improve the efficiency of the PIMD.

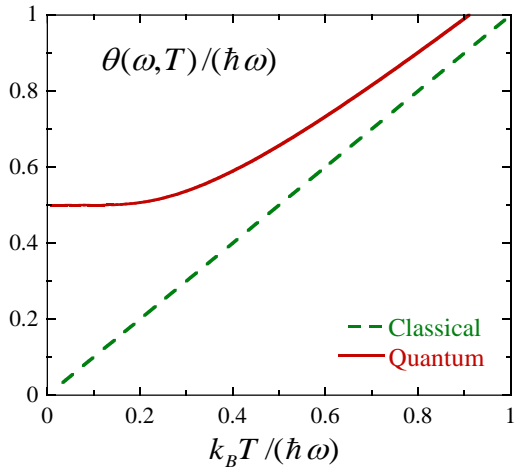

**Figure 3.** Normalized energy of the quantum and classical harmonic oscillators as a function of the reduced temperature.

In the following section, these three methods including NQE in MD simulation will be presented and applied to the problem of ferroelectric phase transitions in BaTiO$_3$ (BTO), taken as a prototypical case. BTO is an anharmonic ferroelectric crystal in which the polar degrees of freedom evolve in a complex multiple-well energy landscape. Quantum effects have been shown to influence its structural properties [8,9]. It undergoes a complex sequence of structural phase transitions [10] as temperature increases: from rhombohedral (R), to orthorhombic (O), tetragonal (T), and cubic (C) structures. The last section will discuss the adequacy of each method.

## 2. Simulation of nuclear quantum effects

### 2.1. Path Integral Molecular Dynamics (PIMD)

NQE can be accounted for by using the path-integral (PI) formalism [4]. In this formulation of quantum statistical mechanics, the canonical partition function $Z$ is written as a discretized imaginary time path integral. For a quantum system containing $N$ (discernible) particles of mass $m$ (assumed as identical for simplicity), $Z$ can be expressed according to:

$$Z = \lim_{P \to \infty} \left(\frac{2\pi m P k_B T}{h^2}\right)^{3NP/2} \times$$
$$\int_{\{\mathbf{r}^N\}^{(1)}} \cdots \int_{\{\mathbf{r}^N\}^{(P)}} \exp\left(-\beta V_{eff}\left(\{r^N\}^{(1)}, \ldots, \{r^N\}^{(P)}\right)\right) \{dr^N\}^{(1)} \cdots \{dr^N\}^{(P)} \quad (2)$$

The integral is over $P$ (Trotter number) replicas of the system, labeled by the integer $s$, each replica being a set of $N$ positions of the atoms $\{r^N\}^{(s)} = \left(\mathbf{r}_1^{(s)}, \ldots, \mathbf{r}_N^{(s)}\right)$. These replicas result from the discretization of the PI in imaginary time (imaginary time slices). The effective potential $V_{eff}$ which depends on all atomic positions of all replicas is composed of two terms, the physical potential energy, $V$, computed in each replica (and averaged over them), and a harmonic coupling term, of angular frequency $\omega_P = \sqrt{P}/\beta\hbar$, between replicas:

$$V_{eff}\left(\{r^N\}^{(1)}, \ldots, \{r^N\}^{(P)}\right) = \sum_{s=1}^{P} \left[\frac{1}{P} V\left(\{r^N\}^{(s)}\right) + \sum_{i=1}^{N} \frac{1}{2} m \omega_P^2 \left(\mathbf{r}_i^{(s)} - \mathbf{r}_i^{(s+1)}\right)^2\right]. \quad (3)$$

Each particle $i$ of the replica $s$ is thus interacting through harmonic forces with the particles $i$ of the replicas $(s+1)$ and $(s-1)$, forming a polymer ring that closes on itself by periodic boundary conditions, $\mathbf{r}_i^{(P+1)} = \mathbf{r}_i^{(1)}$.

In the limit where the Trotter number $P \to \infty$, this equivalent classical system has the same partition function as that of the quantum system. As a consequence, MD simulation can be applied to the classical equivalent to numerically estimate the static properties of the quantum system. In the microcanonical ensemble, the corresponding equation of motion of each particle $i$ in each replica $s$ writes

$$m \ddot{\mathbf{r}}_i^{(s)} = -\frac{1}{P} \nabla_{\mathbf{r}_i^{(s)}} V(\{r^N\}^{(s)}) - m\omega_P^2 \left(2\mathbf{r}_i^{(s)} - \mathbf{r}_i^{(s+1)} - \mathbf{r}_i^{(s-1)}\right). \quad (4)$$

In practical simulations, the Trotter number is finite, and must be chosen to converge the estimated properties. For instance, the average total energy of the system is given by the following estimator:

$$\langle E \rangle = \left\langle \sum_{s=1}^{P} \sum_{i=1}^{N} \frac{\left(\mathbf{p}_i^{(s)}\right)^2}{2m} - \sum_{s=1}^{P} \sum_{i=1}^{N} \frac{1}{2} m\omega_P^2 \left(\mathbf{r}_i^{(s)} - \mathbf{r}_i^{(s+1)}\right)^2 \right\rangle + \left\langle \sum_{s=1}^{P} \frac{1}{P} V(\{r^N\}^{(s)}) \right\rangle. \quad (5)$$

where the first and second averages are the kinetic and the potential energies, respectively. This primitive estimator can be derived from $-\partial Ln(Z)/\partial \beta$.

In the canonical or isothermal-isobaric ensemble, Eq. (4) is modified by adding the forces due the thermostat and/or barostat. In this study of the ferroelectric phase-transitions in BTO, the extension of the Langevin method to the isothermal-isobaric ensemble which has been achieved by Quigley and Probert is used [11,12]. The ferroelectric properties of BTO were modeled by an effective Hamiltonian [13,14] derived from first-principles density-functional calculations. The degrees of freedom of this Hamiltonian are the local modes and the homogeneous strain tensor. The reduced local polar displacement $\mathbf{u}_i$ ($\mathbf{r}_i = a_0 \mathbf{u}_i$) inside the cell $i$ is related to the local dipolar moment $\mathbf{p}_i$ through the cell parameter $a_0$ (7.46 Bohr) and an effective charge $Z^*$ (9.956 $e$): $\mathbf{p}_i = Z^* a_0 \mathbf{u}_i$. Using this model, a ferroelectric phase-transition temperature corresponds to the temperature at which the mean local mode (dipole) rotates and thus exhibits a change of the macroscopic polarization direction: [111], [101] and [001] for the R, O and T phases, respectively.

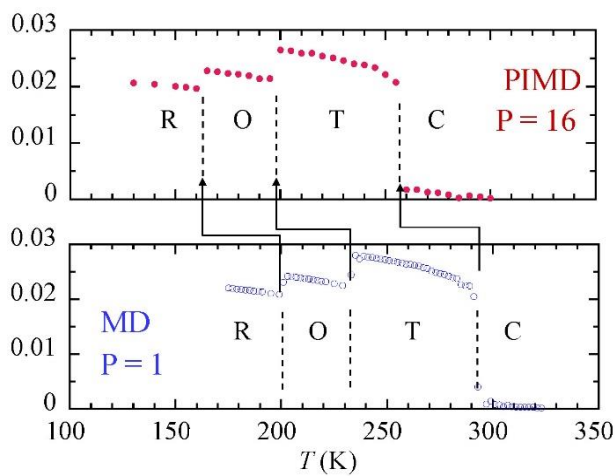

**Figure 4.** Evolution as a function of temperature of the average of the non-zero components of the reduced polarization (local modes) in BaTiO$_3$ as obtained by PIMD and standard MD. Vertical dashed lines show the three ferroelectric phase transitions from rhombohedral (R), to orthorhombic (O), to tetragonal (T), and to cubic (C) structures.

Figure 4 shows the ferroelectric phase-transitions as obtained by PIMD and standard MD. The PIMD calculation is converged with a Trotter number of $P = 16$ in the whole temperature range. The PIMD sequence of phase transitions, R-O-T-C, is identical to the one experimentally observed, but the values of the transition temperatures are different due to the model employed [15]. Hence, these PIMD results will be taken as a reference instead of the experimental ones, in order to test other techniques within the same model. Hence, the standard MD calculation ($P = 1$) overestimates the transition temperatures and the discrepancy is about 30-40 K. This shows an important NQE, which cannot be neglected.

*2.2. Quantum Thermal Bath Molecular Dynamics (QTB-MD)*
An alternative technique to include the NQE in MD simulations is the QTB method [6], which is based on a modification of the Langevin thermostat [16]. The component $\alpha$ of the random force, $R_{i\alpha}$, applied on the atom $i$, is not a *white noise* and its power spectral density, $I_R$, is derived from the quantum dissipation-fluctuation theorem [17], and is related to the Fourier transform of the autocorrelation function, $\langle R_{i\alpha}(t)R_{i\alpha}(t+\tau)\rangle$, according to the Wiener−Khinchin theorem:

$$\langle R_{i\alpha}(t)R_{i\alpha}(t+\tau)\rangle = \int_{-\infty}^{+\infty} I_R(\omega,T) \exp(-i\omega\tau) \frac{d\omega}{2\pi} \tag{6}$$

$$I_R(\omega,T) = 2m\gamma\,\theta(\omega,T) \tag{7}$$

where $\gamma$ is the frictional coefficient. The equation of motion is thus:

$$m\,\ddot{\mathbf{r}}_i = -\boldsymbol{\nabla}_{\mathbf{r}_i} V(\{r^N\}) - m\gamma\dot{\mathbf{r}}_i + \mathbf{R}_i \tag{8}$$

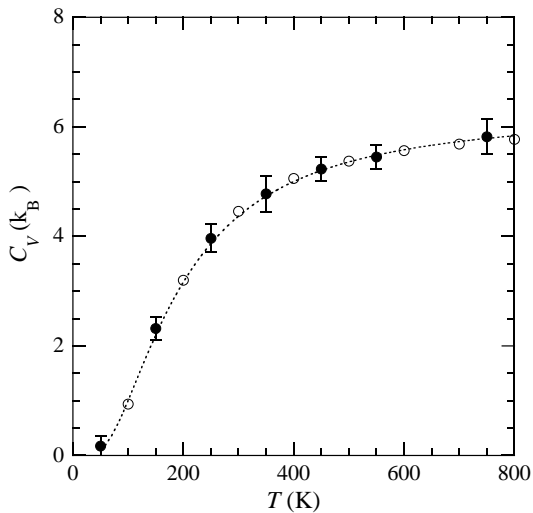

**Figure 5.** Heat capacity of the MgO crystal as computed by QTB-MD [6] (full circles) compared to experimental values (open circles). The dashed line is a guide for the eye.

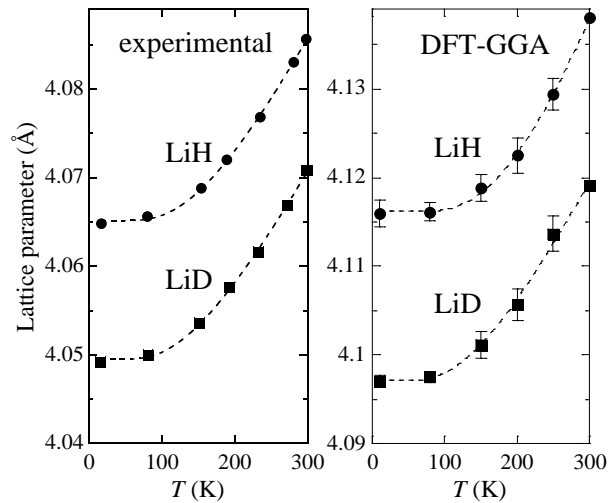

**Figure 6.** Experimental and QTB-MD computation of the lattice parameters of $^7$LiH and $^7$LiD crystals as a function of the temperature, $T$, by using DFT within the GGA [20].

In contrast to the Langevin thermostat, $I_R$ is ω-dependent and the random force components are generated using the procedure detailed in References [18] and [19]. In summary, for $n$ MD time steps, $\delta t$, the random force, $\tilde{R}_{i\alpha}$, is first generated in the Fourier space ($\omega_j = \frac{2\pi j}{n\,\delta t}$):

$$\tilde{R}_{i\alpha}(\omega_j) = (a_j + i\,b_j)\sqrt{I_R(\omega_j)\,n\,\delta t\,/2} \tag{9}$$

where $a_j$ and $b_j$ are normally distributed random numbers, and i the imaginary number. $R_{i\alpha}(t_j = j\delta t)$ is then obtained by inverse Fourier transform.

QTB-MD provides the experimental behavior for the heat capacity in the MgO crystal, as shown in figure 5. The isotopic shift in the lattice parameter of $^7$LiH and $^7$LiD is also well reproduced. Figure 6 yields the experimental and computed lattice parameter at low temperature of both crystals. The discrepancies between experimental and QTB-MD data are due to the GGA functional used. Although exact only in the case of a purely harmonic systems, the QTB leads to satisfactory results in many anharmonic systems, such as MgO [6], LiH [20] and many others [21-24]. Unfortunately, the method can fail when dealing with anharmonic systems. Indeed, the QTB technique is subject to zero-point energy leakage (ZPEL), like any other method based on classical trajectories [25]. It consists in the transfer of energy from high-frequency vibrational modes to low-frequency vibrational modes, hence losing the distribution imposed through the power spectral density of equation (7). A case of ZPEL is given by the ferroelectric phase transitions in $BaTiO_3$. Figure 7a shows that only the T-C phase transition is observed when using usual values of the frictional coefficient − $\gamma$ less than 0.5 THz. The T-C temperature is about 210 K instead of the PIMD reference value of 260 K. In addition, the low-temperature phases, R and O, are not observed even at $T = 0$ K. Fortunately, the ZPEL can be reduced or even suppressed by increasing the value of $\gamma$. As can be seen in figure 7a, the R-O-T-C sequence of transitions occurs from $\gamma = 1.2$ THz. The values of the transition temperatures converge toward the PIMD reference temperatures for higher values of $\gamma$. For $\gamma = 16$ THz, the QTB-MD simulation, with consecutive transition temperatures of 160 K, 190 K and 255 K, is in good agreement with the PIMD result (figure 7b). We conclude that using very high values of the frictional coefficient is an efficient solution in case of failure of the QTB due to ZPEL. The possible drawback of using such high values will be discussed in the last section.

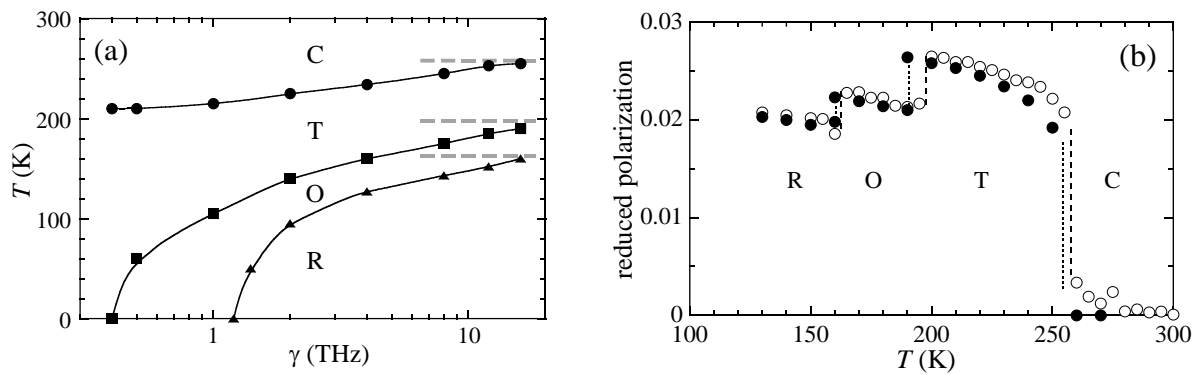

**Figure 7.** R-O, O-T, and T-C ferroelectric phase transitions in $BaTiO_3$. (a) Convergence of the phase-transition temperatures with the friction coefficient, γ, within the QTB-MD simulation. (b) Average of the non-zero components of the reduced polarization (local modes), as a function of the temperature, *T*, obtained by QTB-MD with γ = 16 THz (full circles), and by PIMD with a Trotter number *P* = 16 (open circles).

### 2.3. Combining QTB and PIMD

It is possible to combine the QTB and PIMD [26] in order *i*) to improve the convergence of the PIMD mostly at very low temperature and *ii*) to correct potential failures of the QTB-MD technique especially in the case of anharmonic systems. This combination requires the modification of the power spectral density of the random forces applied on each atom of each replicas. Indeed, for not converged Trotter number, quantum fluctuations are already partially included within the ring polymer of the PIMD. The QTB random forces will thus only bring the missing part of the NQE which is dependent on the Trotter number. In practice, $\theta(\omega,T)$ of equation (7) is replaced by the adequate function $\kappa_P(\omega,T)$, which is solution of the following equation

$$\frac{1}{P}\sum_{k=0}^{P-1}\frac{\kappa_P(\omega_k,T)}{m\,\omega_k^2} = \frac{\theta(\omega,T)}{m\,\omega^2} \tag{10}$$

where $\omega_k$ is the angular frequency of the normal modes of the ring polymer in the harmonic approximation:

$$\omega_k^2 = \frac{\omega^2}{P} + 4\omega_P^2 \sin^2\left(\frac{k\pi}{P}\right) \tag{11}$$

The power spectral density is thus given by:

$$I_R(\omega,T) = 2m\,\gamma\,\kappa_P(\omega,T) \tag{12}$$

Since the corresponding random forces are intended to be applied to the normal modes, the random forces applied on the atoms are obtained through an orthogonal transformation which can be found in Reference [7].

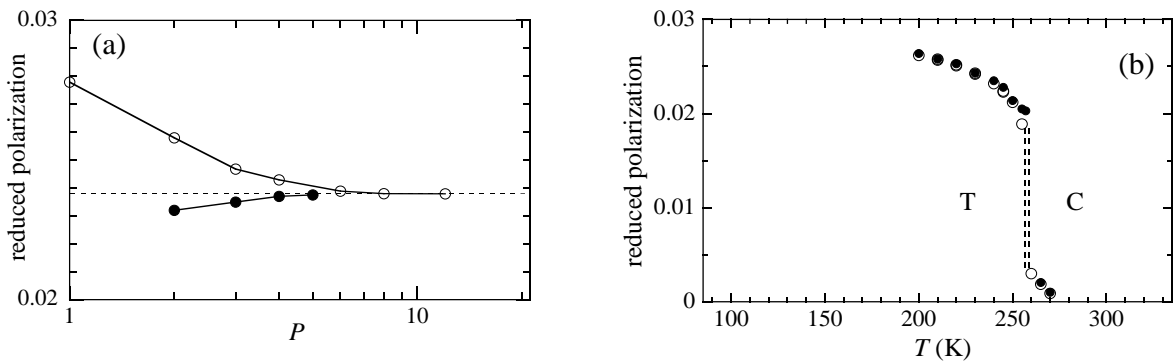

**Figure 8.** T-C ferroelectric phase transitions in BaTiO$_3$. (a) Convergence of the polarization at $T = 240$ K with the Trotter number, $P$, within the QTB-PIMD (full circles), and PIMD (open circles) simulations. (b) Reduced polarization as a function of the temperature, $T$, obtained by QTB-PIMD with $P = 2$ (open circles), and $P = 3$ (full circles).

Let us recall that for QTB-MD simulation, using usual values of the frictional coefficient, the T-C transition in BTO occurs at a temperature of ~210 K, which is much lower than the PIMD reference values (figure 7a). The QTB-PIMD method is applied to the BTO system, varying the Trotter number from 2 to 5 (and using a low value of $\gamma = 0.5$ THz). Figure 8a shows that convergence of the reduced polarization at 240 K in the tetragonal phase is reached for $P = 5$, whereas $P = 8$ is required using PIMD. Hence, the convergence is improved by the combination of the two methods. In figure 8b, QTB-PIMD method with $P = 2$ gives a transition temperature of $T = 257$ K, very close to that obtained by PIMD ($P = 16$, $T = 259$ K). It is worth noting that the failure of the QTB method is fixed solely by two replicas when combining QTB with PIMD. Concerning the R-O and O-T transitions, the transition temperatures are in agreement with the PIMD ones for $P = 3$. This means that the effects of the zero-point energy leakage associated with the use of the QTB have been suppressed by the combination. We conclude, as expected, that QTB-PIMD is more efficient than the standard PIMD, since less replicas are needed for convergence.

## 3. Discussion and Conclusion

The QTB method provides exact results in the case of purely harmonic systems, except for properties that could be affected by frictional forces present in any Langevin dynamics [27]. In addition, the main advantage of the method is that time correlation functions are directly accessible, as in standard MD. Moreover, the QTB is very simple to implement in an existing code.

For anharmonic systems, QTB-MD can fail due to ZPEL, which is the consequence of the coupling between vibrational modes. In this case, the resulting energy distribution does not match equation (1), it is intermediate between the quantum distribution and the classical homogeneous distribution. To suppress or reduce the ZPEL, it is necessary to increase the value of $\gamma$, which is the technical parameter of the thermal bath. Encouraging results have been obtained in the case of a Lennard-Jones aluminum crystal and phase transitions in BTO, as well as for simple models of anharmonic systems [19]. The disadvantage of a high value of $\gamma$ is the broadening of vibrational peaks and a spurious high-frequency tail in the phonon density of states (DOS). Nevertheless, vibrational spectra are not always dramatically altered, this effect depends on the simulated system. It is therefore convenient to check the DOS when using a high value of $\gamma$ within QTB-MD simulations.

Unfortunately, there exist systems for which increasing $\gamma$ do not considerably reduces the ZPEL. The discrimination can be obtained through the strength of the anharmonicity. It can be evaluated with a dimensionless parameter, $C$, including a characteristic length, $d$, and a characteristic energy, $V_0$:

$$C = \frac{\hbar^2}{2mV_0 d^2} \quad (13)$$

There exists a critical value of $C$, depending on the system, below which it can be considered as weakly anharmonic. For the Morse potential and the double-well model the critical values are about 0.01 and 0.1, respectively. Hence, weakly anharmonic systems can be successfully simulated by QTB-MD with an appropriate value of $\gamma$, whereas for strongly anharmonic systems QTB-MD should not be used. In the latter case, QTB-PIMD is an adequate method to take into account the NQE. Indeed, it allows a better convergence with the number of beads than PIMD. The disadvantage of both methods (PIMD and QTB-PIMD) is that time correlation functions are not directly accessible. The sequence of phase transitions in BTO has been successfully retrieved by QTB-MD with a high value of $\gamma$ [19] and by QTB-PIMD [7]. In contrast, the values of the ferroelectric polarization in BTO close to the temperatures of the phase-transitions are better computed in QTB-PIMD.

For weakly anharmonic systems, some macroscopic properties are not affected by the ZPEL. In fact, the total energy is exact although the energy distribution is not correct. For instance, QTB-MD has been successfully used in numerous applications to highlight nuclear quantum effects or isotope effect on lattice parameter [6,20], on heat capacity of crystal [6] and carbon nanotube [22], and on binding energies in clusters [28].